\documentclass[
preprint,
showpacs,preprintnumbers,
bibnotes,
 amsmath,amssymb,
 aps,
 pra,
superscriptaddress,
longbibliography,
]{revtex4-1}

\usepackage{amsmath}
\usepackage{amsfonts}
\usepackage{amssymb}
\usepackage{graphicx}
\usepackage{xspace}
\usepackage{color}
\usepackage{comment}
\usepackage[hidelinks]{hyperref}
\usepackage{dcolumn}
\usepackage{bm}
\usepackage{mathtools}

\usepackage{comment}

\usepackage[detect-weight=true, detect-family=true]{siunitx}

\graphicspath{ {./figures/} }

\begin{document}

\title{
Low-threshold exciton transport and control in atomically thin semiconductors
}
\author{Hyeongwoo Lee$^{\dagger}$}
\affiliation
{Department of Physics, Ulsan National Institute of Science and Technology (UNIST), \\Ulsan 44919, Republic of Korea}
\author{Yeonjeong Koo$^{\dagger}$}
\affiliation
{Department of Physics, Ulsan National Institute of Science and Technology (UNIST), \\Ulsan 44919, Republic of Korea}
\author{Jinseong Choi}
\affiliation
{Department of Physics, Ulsan National Institute of Science and Technology (UNIST), \\Ulsan 44919, Republic of Korea}
\author{Shailabh Kumar}
\affiliation
{Department of Medical Engineering, California Institute of Technology (Caltech), \\CA 91125, USA}
\author{Hyoung-Taek Lee}
\affiliation
{Department of Physics, Ulsan National Institute of Science and Technology (UNIST), \\Ulsan 44919, Republic of Korea}
\author{Gangseon Ji}
\affiliation
{Department of Physics, Ulsan National Institute of Science and Technology (UNIST), \\Ulsan 44919, Republic of Korea}
\author{Soo Ho Choi}
\affiliation
{Center for Integrated Nanostructure Physics, Institute for Basic Science (IBS), \\Suwon 16419, Republic of Korea}
\author{Mingu Kang}
\affiliation
{Department of Physics, Ulsan National Institute of Science and Technology (UNIST), \\Ulsan 44919, Republic of Korea}
\author{Ki Kang Kim}
\affiliation
{Center for Integrated Nanostructure Physics, Institute for Basic Science (IBS), \\Suwon 16419, Republic of Korea}
\affiliation
{Department of Energy Science, Sungkyunkwan University (SKKU), \\Suwon 16419, Republic of Korea}
\author{Hyeong-Ryeol Park}
\affiliation
{Department of Physics, Ulsan National Institute of Science and Technology (UNIST), \\Ulsan 44919, Republic of Korea}
\author{Hyuck Choo}
\affiliation
{Department of Medical Engineering, California Institute of Technology (Caltech), \\CA 91125, USA}
\affiliation
{Imaging Device Lab, Device $\&$ System Research Center, Samsung Advanced Institute of Technology (SAIT), \\Suwon 16678, Korea}
\author{Kyoung-Duck Park$^{\ast}$}
\affiliation
{Department of Physics, Ulsan National Institute of Science and Technology (UNIST), \\Ulsan 44919, Republic of Korea}
\email{kdpark@unist.ac.kr}
\date{\today}

\begin{abstract}

\noindent 
\textbf{
Understanding and controlling the nanoscale transport of excitonic quasiparticles in atomically thin 2D semiconductors is crucial to produce highly efficient nano-excitonic devices.
Here, we present a nano-gap device to selectively confine excitons or trions of 2D transition metal dichalcogenides at the nanoscale, facilitated by the drift-dominant exciton funnelling into the strain-induced local spot.
We investigate the spatio-spectral characteristics of the funnelled excitons in a WSe$_2$ monolayer (ML) and converted trions in a MoS$_2$ ML using hyperspectral tip-enhanced photoluminescence (TEPL) imaging with $<$15 nm spatial resolution.
In addition, we dynamically control the exciton funnelling and trion conversion rate by the GPa scale tip pressure engineering.
Through a drift-diffusion model, we confirm an exciton funnelling efficiency of $\sim$25 $\%$ with a significantly low strain threshold ($\sim$0.1 $\%$) which sufficiently exceeds the efficiency of $\sim$3 $\%$ in previous studies.
This work provides a new strategy to facilitate efficient exciton transport and trion conversion of 2D semiconductor devices.
}

\end{abstract}

\maketitle

\noindent 
A distinguished advantage of atomically thin semiconductors compared to other types of low-dimensional quantum materials, such as 0D quantum dots and 1D nanowires, is the wide exciton distribution over a 2D area.
On the one hand, this fascinating property provides significant benefits to fabricate highly efficient and ultrathin optoelectronic devices operating at room temperature, especially in photovoltaics \cite{bernardi2013} and light emitting devices \cite{ross2014}.
On the other hand, to exploit excitons as carriers for quantum information devices and exciton integrated circuits, e.g., using single-photon emitting localized excitons \cite{lee2021afm} or long coherence dark excitons \cite{park2018} as well as neutral- and charged-excitons \cite{moon2020, harats2020}, controlling the exciton dynamics is a challenging subject.
In order to manipulate the exciton dynamics of 2D transition metal dichalcogenides (TMDs), e.g., drift-induced excitonic flux and diffusion-induced energy conversion, various strain-engineering approaches have been demonstrated recently \cite{moon2020, harats2020}.
The electronic band structure and excitonic properties of 2D semiconductors can be significantly modified by tuning the crystal strain owing to their atomic thickness.
Hence, many studies attempted to achieve the deterministic exciton funnelling to a low bandgap region by fabricating strain-gradient devices, e.g., using wrinkled substrates \cite{lee2021}, piezoelectric actuators \cite{haque2003}, and atomic force microscopy (AFM) tips \cite{moon2020, harats2020}.
However, recent works revealed that the exciton funnelling efficiency using these approaches is significantly low ($<$3 $\%$) at room temperature, due to the dominant diffusion in the microscale bandgap-gradient regions overwhelming the required drift process \cite{feng2012, harats2020}.
Thus, most studies induced a massive strain on the crystal, deteriorating the crystal quality and radiative quantum yield \cite{lee2021}.
Therefore, designing an ideal TMD device, e.g., with high exciton funnelling efficiency and large quantum yield with an insignificant induced strain, is highly desirable.
In addition, to raise the degree of integration of quantum devices and exciton integrated circuits, the exciton funneling channel must be significantly scaled down compared to the current microscale dimensions.

Another intriguing observation in the previous study, was the trion conversion at the funnelling region in a WS$_2$ monolayer (ML) \cite{harats2020}.
In particular, they obtained that the free carriers are efficiently funnelled into the lowest bandgap region and coupled to neutral excitons (X$_0$) to form the trion (X-) state with almost 100 $\%$ conversion efficiency without electrical gating even at room temperature.
This remarkable property can widely expand the scope of applications in 2D excitonic devices with the ability to control the dynamics of excitons and trions.
However, because the previous work only investigated the exciton funnelling and trion conversion of n-type TMD MLs with a microscale strain-gradient with diffraction-limited far-field spectroscopy, it introduced confusion with the role of doping types and the size of the bandgap-gradient region.
Hence, a high resolution spatio-spectral investigation for various TMD MLs with a correlated analysis of the structural, optical, and theoretical properties is crucial to understand the exciton dynamics at its natural length scale comprehensively.

Here, we present a nano-gap device to facilitate the low-threshold and highly efficient exciton funnelling and trion conversion processes even at room temperature.
The nanoscale strain-gradient region induced by the nano-gap enables the drift-dominant region to occupy over $\sim$60 $\%$ of the strain-gradient area.
It leads to the high exciton funnelling efficiency exceeding $\sim$25 $\%$ estimated by a drift-diffusion model designing the WSe$_2$ and MoS$_2$ MLs on the nano-gap \cite{harats2020}.
To experimentally confirm these nanoscale exciton funnelling and trion conversion, we use hyperspectral tip-enhanced photoluminescence (TEPL) imaging with a spatial resolution of $\sim$15 nm (see Fig. S1 in Supplementary Information for details).
It reveals that the observed PL enhancement of $\sim$180 $\%$ with $\sim$0.1 $\%$ strain is comparable to the PL enhancement from the microscale strain-gradient achieved with over $\sim$10 $\%$ strain for a WSe$_2$ ML \cite{lee2021}.
For a MoS$_2$ ML, in contrast, exciton to trion conversion is observed at the nano-gap region. 
In addition, we utilize the plasmonic tip for nano-opto-mechanical control of these TMD crystals.
By directly pressing and releasing the crystal with GPa scale tip pressure, we dynamically engineer $\sim$10 times greater strain on the TMD crystal at the nanoscale. 
This experiment demonstrates the ability to gradually tune the exciton funnelling and trion conversion rates in TMD MLs in a reversible manner.
\\

\begin{figure*}
	\includegraphics[width = 16 cm]{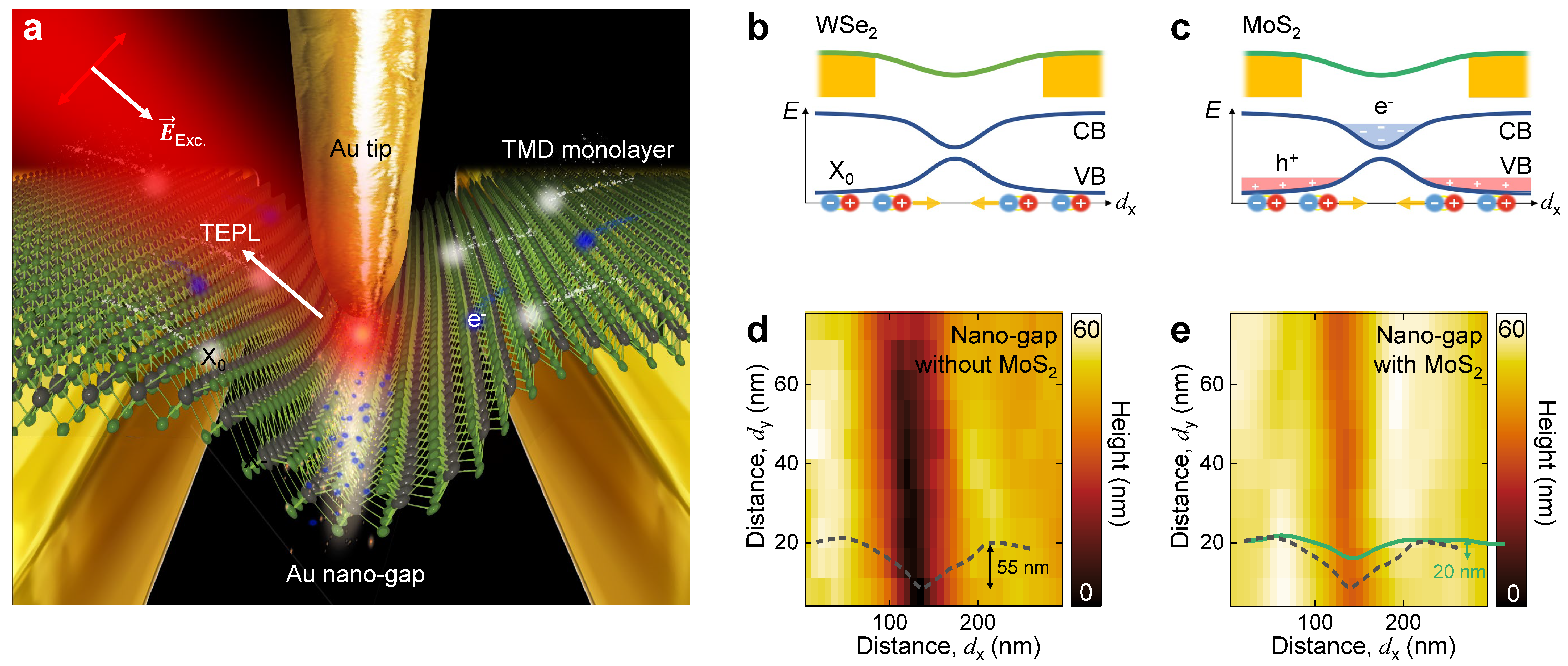}
	\caption{
\textbf{Schematic illustrations of an experimental design and energy diagram.} 
(a) The Au nano-gap device and transferred TMD MLs combined with TEPL spectroscopy to probe and control the electric charges (e$^-$) and exciton (X$_0$) dynamics at the nanoscale. 
Illustrations of the energy band diagram for WSe$_2$ (b) and MoS$_2$ (c) MLs on the nano-gap and spatial distributions of electric charges (e$^-$ and h$^+$) and photoexcited excitons (X$_0$). 
AFM topography images and height profiles of the nano-gap without (d) and with (e) MoS$_2$ ML exhibiting a wrinkled crystal structure which gives rise to a nanoscale strain-gradient. 
}
	\label{fig:fig1}
\end{figure*}

\noindent
\textbf{Pre-characterizations of TMD MLs on the nano-gap}

\noindent
To facilitate an efficient exciton funnelling and trion conversion for WSe$_2$ and MoS$_2$ MLs and further investigate their spatial dynamics in the sub-wavelength scale region, we use the nano-gap device combined with TEPL spectroscopy, as illustrated in Fig.~\ref{fig:fig1}a.
The WSe$_2$ and MoS$_2$ MLs are transferred onto the nano-gap device with a gap size of $\sim$150-300 nm, to induce nanoscale strain-gradient to the MLs crystals at the nano-gap region (see Methods for more details).
When the excitation beam (632.8 nm) is focused to this TMD MLs on the nano-gap, the photo-excited excitons are generated and drifted toward the nano-gap center, i.e., the lowest potential energy region, attributed to the strain-gradient effect. 
Fig.~\ref{fig:fig1}b and 1c show energy diagrams of the WSe$_2$ and MoS$_2$ MLs on the nano-gap describing the distinguished dynamics of electric charges and excitons in the same strain-gradient condition.
The largest strain induced at the gap center gives rise to the smallest bandgap for both WSe$_2$ and MoS$_2$ MLs, which stimulates the funnelling of excitons and electrons toward it.
In the WSe$_2$ ML, only neutral excitons are dominantly funnelled toward the gap because it has few electrons. 
In contrast, in the MoS$_2$ ML, electrons are dominantly drifted to the gap apart from the exciton funnelling \cite{harats2020} because it is naturally n-type doped material \cite{singh2019}. 
Additionally, the funnelled excitons are converted to the trion state by the overflowing electrons in the gap.
We use TEPL spectroscopy to characterize these exciton dynamics and trion conversion in their natural physical length scale, as shown in Fig.~\ref{fig:fig1}a.
Note that we maintain $\sim$10 nm tip-sample distance to avoid the possible perturbation effect to the crystal by the Au tip.

In addition to probing exciton and trion properties, this approach controls these dynamics at the nanoscale, as shown in Fig.~\ref{fig:fig4}.
Specifically, in our home-built TEPL setup, the electrochemically etched Au tip (apex radius of $\sim$15 nm) enhances PL signal and gives a spatial resolution of $\sim$15 nm by the nano-optical antenna effect and Purcell enhancement \cite{lee2021acs, lee2021afm}. 
Furthermore, our shear-force AFM module offers a nanoscale 3D positioning of the Au tip with $<$0.2 nm precision, facilitating a systematic strain engineering and exciton/trion control of the WSe$_2$ and MoS$_2$ MLs with measuring TEPL signals simultaneously (see Methods for more details).
The well-suspended TMD ML on the nano-gap to generate the strain-gradient effect is pre-characterized through AFM measurements, as shown in Fig.~\ref{fig:fig1}d and 1e.
Topography images of the nano-gap with and without the MoS$_2$ ML exhibit a difference in depth, as indicated by the topographic line profiles in Fig.~\ref{fig:fig1}e.
\\

\begin{figure*}
	\includegraphics[width = 16 cm]{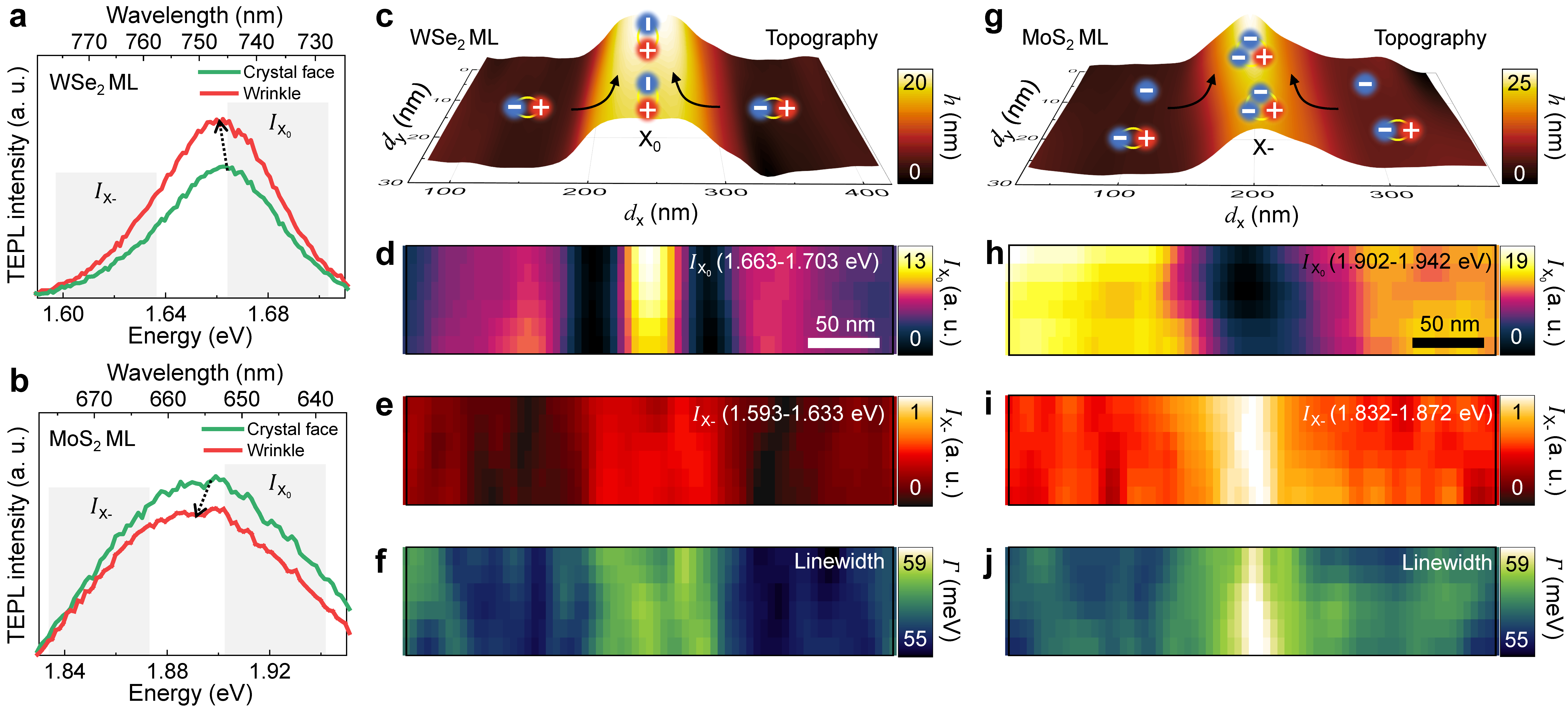}
	\caption{
\textbf{Hyperspectral TEPL imaging of strained TMD MLs at the wrinkle.} 
TEPL spectra of WSe$_2$ (a) and MoS$_2$ (b) MLs at the crystal face (green) and the wrinkle (red) regions.
(c-f) Hyperspectral TEPL images of a WSe$_2$ ML. AFM topography image with a description of exciton funnelling (c). TEPL images of the spectrally integrated intensity of excitons (d, spectral region of \textit{I$_{X_0}$} in Fig.~\ref{fig:fig2}a) and low energy shoulder (e, spectral region of \textit{I$_{X-}$} in Fig.~\ref{fig:fig2}a after normalization). TEPL image of spectral linewidth (f).
(g-j) Hyperspectral TEPL images of a MoS$_2$ ML. AFM topography image with a description of electron funnelling and trion (X-) conversion (g). TEPL images of the spectrally integrated intensity of excitons (h, spectral region of \textit{I$_{X_0}$} in Fig.~\ref{fig:fig2}b) and low energy shoulder (i, spectral region of \textit{I$_{X-}$} in Fig.~\ref{fig:fig2}b after normalization). TEPL image of spectral linewidth (j). 
}
	\label{fig:fig2}
\end{figure*}

\noindent
\textbf{Exciton dynamics and trion conversions in nanoscale wrinkles}

\noindent
The ideal nanoscale strain-gradient structure for effective X$_0$ funnelling and X- conversion exists in nature, i.e., the nanoscale wrinkle in 2D TMDs.
Our previous study revealed the exciton funnelling behaviors of the naturally-formed wrinkles in a WSe$_2$ ML \cite{koo2021}.
As experimentally confirmed again in this study, the uniaxial tensile strain is induced in the wrinkle structure with the highest strain at the apex region, which facilitates the exciton funnelling to the lowest bandgap energy region, spatially at the nanoscale.
Fig.~\ref{fig:fig2}a shows TEPL spectra of a WSe$_2$ ML when the Au tip is placed at the crystal face and the wrinkle. 
The peak redshift of $\sim$3 meV at the wrinkle compared to the peak energy at the crystal face allows us to estimate the induced tensile strain of $\sim$0.05 $\%$ at the wrinkle apex \cite{schmidt2016}.
The TEPL intensity at the wrinkle increases $\sim$140 $\%$ compared to the TEPL intensity at crystal face by the exciton funnelling effect.
This PL increase ratio is comparable to the previous study obtained by inducing a significantly higher strain of $\sim$5 $\%$ with the microscale strain-engineering technique \cite{lee2021}. 
That is a dramatically improved exciton funnelling efficiency with respect to the induced strain is obtained at this nanoscale strain-gradient structure (wrinkle).
The specific details for the physical mechanism are depicted with a theoretical model in Fig.~\ref{fig:fig5}, which was not considered in our recent report \cite{koo2021}.
In contrast, the TEPL intensity decreases with distinctly increased linewidth at the wrinkle for a MoS$_2$ ML compared to the crystal face region, as shown in Fig.~\ref{fig:fig2}b.
This inverse feature at the wrinkle between WSe$_2$ and MoS$_2$ MLs is attributed to the doping type of TMD crystals.
Therefore, the electron funnelling along with a trion conversion is the dominant process at the wrinkle for the n-type doped MoS$_2$ ML.
We will discuss the spatial dynamics of excitons and electric charges, and the trion conversion, for different TMDs in the nanoscale strain-gradient structures after demonstrating the results of our control experiments first.

To investigate the efficient exciton funnelling in a WSe$_2$ ML and the trion conversion in a MoS$_2$ ML, we obtain hyperspectral TEPL images at the wrinkles (Fig.~\ref{fig:fig2}c-j).
As expected from the TEPL point spectra in Fig.~\ref{fig:fig2}a, a large TEPL signal at the wrinkle apex is observed in the integrated intensity image for neutral excitons (\textit{I$_{X_0}$}, 1.663-1.703 eV, Fig.~\ref{fig:fig2}c-d) for a WSe$_2$ ML.
We also observe the marginally increased integrated intensity for trions (\textit{I$_{X-}$}, 1.593-1.633 eV, Fig.~\ref{fig:fig2}e) with the broaden linewidth (Fig.~\ref{fig:fig2}f). 
However, the significant modifications are not observed compared to the X$_0$ emission.  
On the contrary, the TEPL intensity for neutral excitons (\textit{I$_{X_0}$}, 1.902-1.942 eV) shows a pronounced decrease at the wrinkle for a MoS$_2$ ML, as shown in Fig.~\ref{fig:fig2}g-h. 
Furthermore, we observe the distinct increase of the integrated intensity for trions (\textit{I$_{X-}$}, 1.832-1.872 eV, Fig.~\ref{fig:fig2}i) with the significantly broaden linewidth (Fig.~\ref{fig:fig2}j) at the wrinkle apex because of the efficient trion conversion.
\\

\begin{figure*}
	\includegraphics[width = 16 cm]{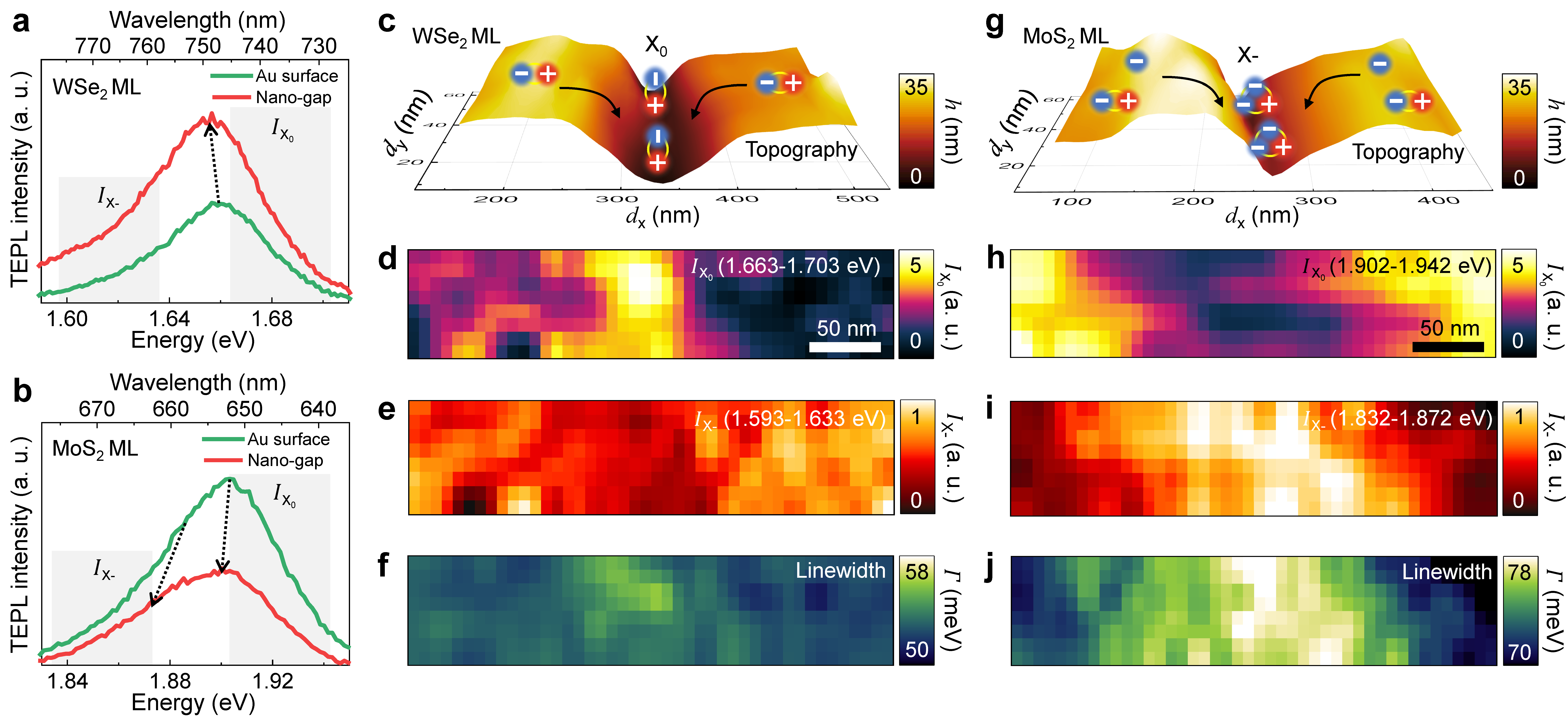}
	\caption{
\textbf{Hyperspectral TEPL imaging of TMD MLs at the nano-gap.} 
TEPL spectra of WSe$_2$ (a) and MoS$_2$ (b) MLs at the Au surface (green) and the nano-gap (red) regions.
(c-f) Hyperspectral TEPL images of a WSe$_2$ ML. AFM topography image with a description of exciton funnelling (c). TEPL images of the spectrally integrated intensity of excitons (d, spectral region of \textit{I$_{X_0}$} in Fig.~\ref{fig:fig3}a) and low energy shoulder (e, spectral region of \textit{I$_{X-}$} after normalization in Fig.~\ref{fig:fig3}a after normalization). TEPL image of spectral linewidth (f).
(g-j) Hyperspectral TEPL images of a MoS$_2$ ML. AFM topography image with a description of electron funnelling and trion (X-) conversion (g). TEPL images of the spectrally integrated intensity of excitons (h, spectral region of \textit{I$_{X_0}$} in Fig.~\ref{fig:fig3}b) and low energy shoulder (i, spectral region of \textit{I$_{X-}$} in Fig.~\ref{fig:fig3}b after normalization). TEPL image of spectral linewidth (j).
}
	\label{fig:fig3}
\end{figure*}

\noindent
\textbf{Deterministic exciton funnelling and trion conversion using nano-gap}

\noindent
Nanoscale wrinkles are attractive nanostructures for efficient exciton funnelling and trion conversion.
However, their applications to exciton integrated circuits or other excitonic devices are restricted due to their random locations.
Hence, to deterministically control the exciton energy and spatial dynamics at the nanoscale, we fabricate the strain-gradient of WSe$_2$ and MoS$_2$ MLs by the nano-gap, i.e., the inverse-wrinkle structure at a specific location.
Fig.~\ref{fig:fig3}a shows a TEPL spectrum of a WSe$_2$ ML at the nano-gap exhibiting intensity increase with slight peak redshift (red), compared to the TEPL spectrum measured at the Au surface (green).
The peak redshift is attributed to the bandgap bending induced by a uniaxial tensile strain of $\sim$0.1 $\%$ \cite{schmidt2016}, which is followed by the exciton funnelling and consequent TEPL intensity increase of $\sim$180 $\%$.
These features are similar to the results of a WSe$_2$ wrinkle (Fig.~\ref{fig:fig2}a).
In contrast, Fig.~\ref{fig:fig3}b shows an asymmetric TEPL spectrum of a MoS$_2$ ML on the Au surface by the existence of trions (green).
The asymmetricity of a TEPL spectrum measured at the nano-gap becomes significantly apparent with decreased emission intensity.
This different spectral feature between the WSe$_2$ and MoS$_2$ crystals is originated from the excess electrons in the MoS$_2$ ML.
Because MoS$_2$ is an electron-rich TMD crystal, the distinguishable X- shoulder is naturally observed in the flat crystal even at room temperature \cite{singh2019, dolui2013}.
The X- peak becomes more significant with the local strain-gradient because the electrons are funnelled to the lower bandgap energy region together with neutral excitons and they are converted to trions \cite{harats2020}. 
The decreased X$_0$ TEPL intensity with the emerging X- peak at the nano-gap center well supports this trion conversion process.

To more clearly investigate their exciton funnelling and trion conversion processes, we perform hyperspectral TEPL imaging.
Fig.~\ref{fig:fig3}c-f show the measured AFM topography, integrated TEPL intensity images of neutral excitons (\textit{I$_{X_0}$}, 1.663 - 1.703 eV) and trions (\textit{I$_{X-}$}, 1.593-1.633 eV), and TEPL linewidth map of a WSe$_2$ ML.
We observe the expected spatial dynamics of excitons moving into the topographical valley, i.e., exciton drifts toward the strain-maximum region. 
In contrast, the spatial maps of trion TEPL intensity and TEPL linewidth show no significant differences at the nano-gap compared to the Au surface region.
The MoS$_2$ ML on the nano-gap shows different behaviors for neutral excitons (\textit{I$_{X_0}$}, 1.902-1.942 eV) and trions (\textit{I$_{X-}$}, 1.832-1.872 eV), as shown in Fig.~\ref{fig:fig3}g-j. 
Because the drifted neutral excitons and electrons are expected to be converted to trions, \textit{I$_{X_0}$} decreases at the nano-gap (Fig.~\ref{fig:fig3}h).
Meanwhile, the \textit{I$_{X-}$} shows a considerable increase at the nano-gap (Fig.~\ref{fig:fig3}i) and the converted trion peak gives rise to the linewidth increase (Fig.~\ref{fig:fig3}j).
Our artificially fabricated strain-gradient TMDs at the nano-gap generally show similar characteristics to the naturally-formed nanoscale wrinkles in WSe$_2$ and MoS$_2$ MLs.
Therefore, we confirm the on-demand capability of deterministic exciton funnelling and trion conversion at the nanoscale with high efficiency.
\\

\begin{figure*}
	\includegraphics[width = 16 cm]{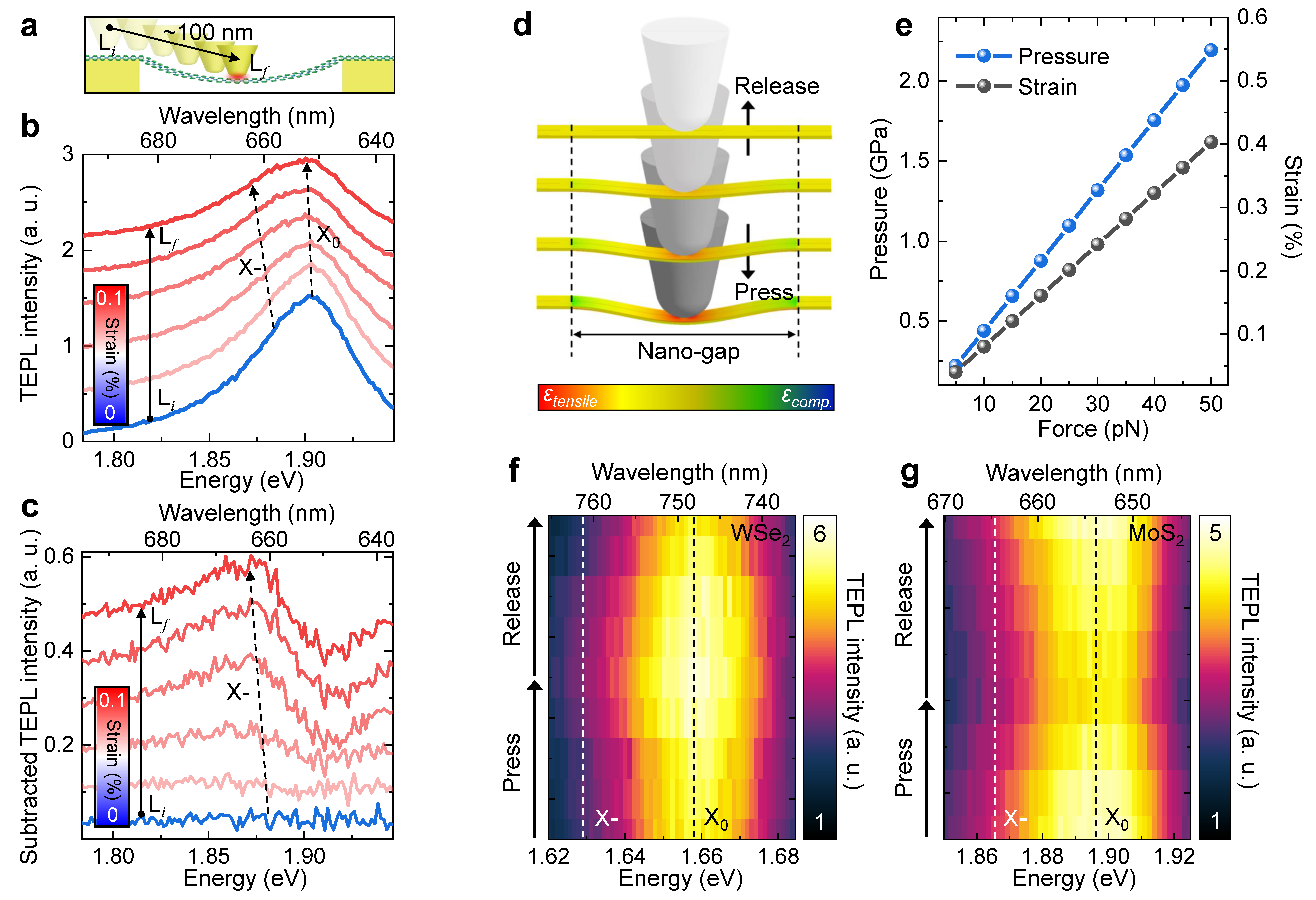}
	\caption{
\textbf{Tip-induced control of exciton funnelling and trion conversion.} 
(a) Schematic illustration for the lateral tip movement from L$_i$ to L$_f$ (distance of $\sim$100 nm). 
(b) TEPL spectra of a MoS$_2$ ML at the nano-gap when the Au tip moves from L$_i$ to L$_f$. 
(c) Subtracted TEPL spectra exhibiting the pronounced spectral modification. In order to obtain these results, all the spectra in Fig.~\ref{fig:fig4}b are normalized and subtracted by the spectrum at 0 $\%$ strain (blue in Fig.~\ref{fig:fig4}b).
(d) Simulated strain distribution for a TMD MLs on the nano-gap as the Au tip presses and releases the crystal. 
(e) Calculated local pressure and strain of the crystal as a function of the force applied by the Au tip. 
(f, g) Reversibly engineered TEPL spectra of WSe$_2$ (f) and MoS$_2$ (g) MLs by the tip force, exhibiting nanoscale dynamics of excitons and trions. Black and white dashed lines indicate the energy of neutral excitons (X$_0$) and trions (X-).  
}
	\label{fig:fig4}
\end{figure*}

\noindent
\textbf{Dynamic control of exciton funnelling and trion conversion via GPa tip-pressure}

\noindent 
We investigate the detailed trion conversion process of a MoS$_2$ ML at the nanoscale strain-gradient by measuring the evolution of TEPL spectra when the Au tip laterally moves toward the nano-gap center (Fig.~\ref{fig:fig4}a).
As shown in Fig.~\ref{fig:fig4}b, the gradual redshift of X$_0$ peak represents the formation of strain-gradient of a MoS$_2$ ML, which has a maximum strain of $\sim$0.1 $\%$ at the center \cite{conley2013}. 
In addition, the X- shoulder emerges, and its intensity increases as the induced strain increases. 
To better represent the spectral changes with respect to the increased strain, we normalize the series of TEPL spectra and subtract the TEPL spectrum measured at the Au surface from them, as shown in Fig.~\ref{fig:fig3}c.
The X- peak evolution from the edge to the nano-gap center is observed in the subtracted TEPL spectra.

We then demonstrate dynamic control of exciton funnelling and trion conversion by directly pressing and releasing the crystal with the Au tip. 
In our numerical simulations (Fig.~\ref{fig:fig4}d), the strain-gradient maximum shows the gradual increase as the Au tip presses a TMD ML at the nano-gap center.
Accordingly, we can derive the applied pressure and the strain on a TMD ML as a function of the applying force by the Au tip, as shown in Fig.~\ref{fig:fig4}e.
This result shows that the key feature of our approach is the control of extremely large pressure at the GPa scale with the $\sim$pN scale force because the unit area of Au tip is on the order of 10$^2$ square nanometers.
In addition, we can simultaneously measure TEPL responses with spatial resolution of $\sim$15 nm and dynamically engineer the tip-induced strain on the crystal, which was not allowed in other works \cite{darlington2020}.
Fig.~\ref{fig:fig4}f and g show the experimental results of tip-induced strain control and corresponding modification of TEPL spectra for WSe$_2$ and MoS$_2$ MLs.
When the Au tip presses the WSe$_2$ ML, X$_0$ TEPL intensity increases with the increasing exciton funnelling rate whereas it decreases to the original state when the tip releases from the crystal. 
In contrast, X$_0$ TEPL intensity decreases as the Au tip presses the MoS$_2$ ML with the increasing trion conversion rate whereas it returns to the original state as the tip releases.
The detailed analysis on these reversible TEPL intensity changes of X$_0$ and X- is presented in Fig. S2.
\\

\begin{figure*}
	\includegraphics[width = 16 cm]{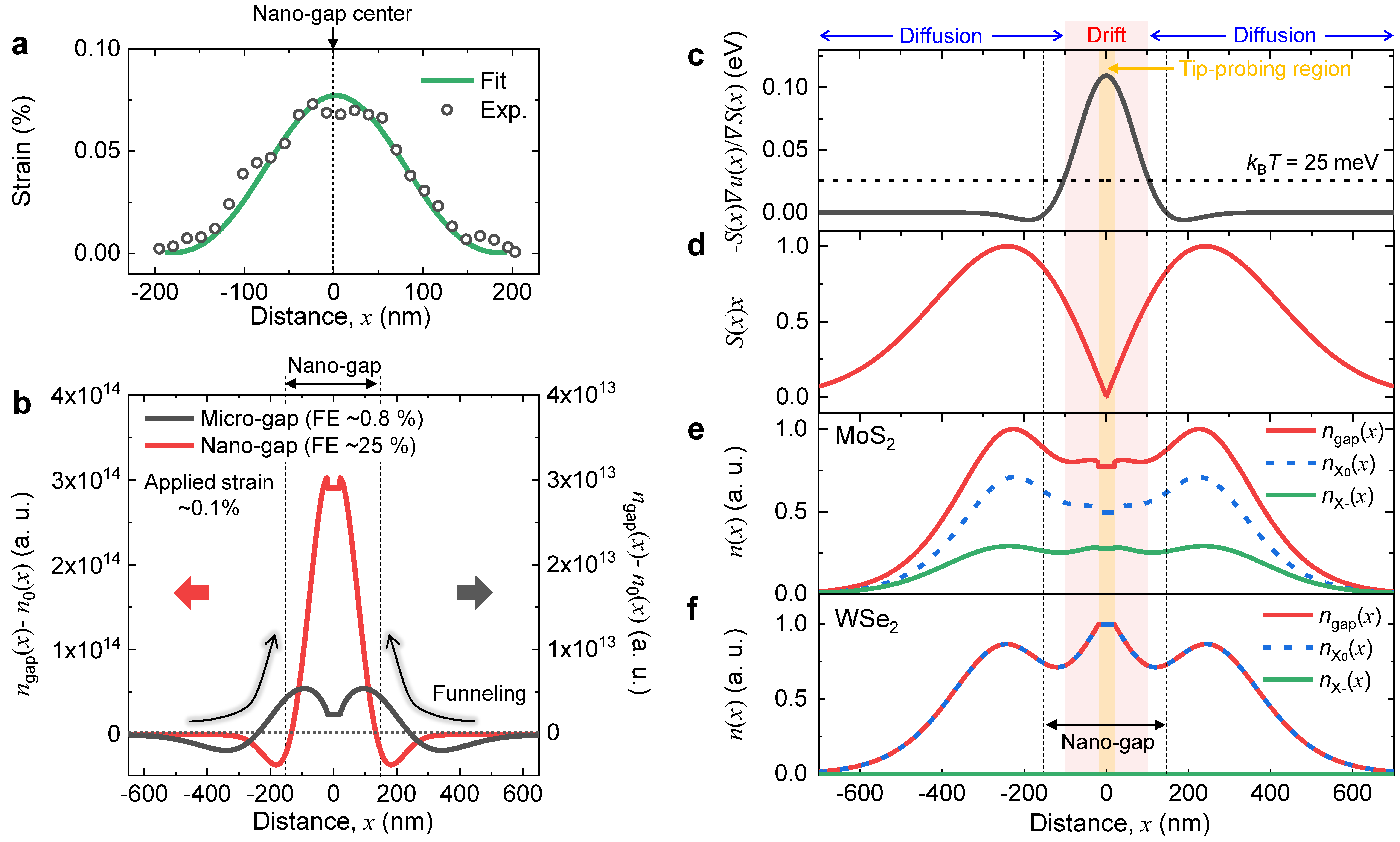}
	\caption{
		\textbf{Spatial distribution of excitons and trions at the nano-gap.} 
(a) Strain distribution of a MoS$_2$ ML on the nano-gap obtained from exciton TEPL peak energy shift (black dots) and fitted line shape function (green line). 
(b) Spatial density distribution of neutral excitons at the micro-gap (black) and the nano-gap (red) for a MoS$_2$ ML under the applied strain of $\sim$0.1 $\%$. \textit{n$_{gap}$(x)} and \textit{n$_{0}$(x)} indicate their spatial density with and without the gap, respectively. 
(c) Spatial distribution of the ratio between the drift and diffusion terms (-$\frac{S(x){\nabla}u(x)}{{\nabla}S(x)}$, solid line) and thermal energy \textit{k$_{B}$T} at room temperature (dashed line). The drift dominant (shaded) and diffusion dominant regions are classified. The yellow shaded region indicates the tip-probing region in TEPL measurements.
(d) Spatial distribution of the optical source term \textit{S(x)x} in our experiment.
(e, f) Normalized density distribution of neutral excitons \textit{n$_{X_0}$(x)} and trions \textit{n$_{X-}$(x)} at the nano-gap for MoS$_2$ (e) and WSe$_2$ (f) MLs.}
	\label{fig:fig5}
\end{figure*}

\noindent
\textbf{Theoretical investigations of nanoscale exciton transport}

\noindent
To quantitatively understand the advantages of the nanoscale strain-gradient compared to the previously studied microscale strain-gradient geometry \cite{moon2021, harats2020, lee2021}, we perform numerical analysis on TMD MLs under the strain-gradient. 
First, we model a strain-gradient profile based on the experimentally obtained strain distribution at the nano-gap, as shown in Fig.~\ref{fig:fig5}a (see Fig. S3 for more details) \cite{conley2013}.
We then analyse the drift-diffusion equation of MoS$_2$ and WSe$_2$ MLs with the modeled strain-gradient profile \cite{harats2020, kulig2018}. 
The experimental values of Auger recombination rate, exciton lifetime, and diffusion coefficient of WSe$_2$ and MoS$_2$ MLs for solving the equation are derived from previous studies \cite{uddin2020, konabe2014, palummo2015}.
Fig.~\ref{fig:fig5}b shows the exciton density distributions of a MoS$_2$ ML at the $\sim$3 $\mu$m micro-gap and the $\sim$300 nm nano-gap under the same applied strain of $\sim$0.1 $\%$.
To investigate the contribution of exciton drift to the spatial distribution, we subtract the exciton density without strain-gradient \textit{n$_0$(x)} from the exciton density with strain-gradient \textit{n$_{gap}$(x)} (see Fig. S3 for more details).
For both micro-gap and nano-gap, the exciton density increases at the center of the gap whereas it decreases in the vicinity, representing the exciton funnelling phenomenon. 
However, the concentration of the drifted-exciton at the center is much weaker and broader at the micro-gap compared to the nano-gap.
To quantitatively estimate the exciton funnelling efficiency of the micro-gap and nano-gap to the $\sim$15 nm tip apex region, we define the following equation$:$
\begin{equation} 
\begin{array}{cl}
funnelling$\;$efficiency, \textit{$\eta$} = \cfrac{{\int}_{0}^{x_{tip}}\{n_{gap}(x)-n_{0}(x)\}dx}{{\int}_{0}^{x_{gap}}\{n_{gap}(x)-n_{0}(x)\}dx}$\;$.
\end{array}
\label{eq1}
\end{equation}
We obtain \textit{$\eta$} $\approx$ 25 $\%$ and 0.8 $\%$ for the nano- and micro-gaps, respectively (see calculation of the funnelling efficiency in Supplementary Information for more details).
Note that we exclude the existence of the electron in this step to precisely investigate the exciton funnelling efficiency.

We can understand the dramatically enhanced funnelling efficiency at the nano-gap by comparing the portion of drift-dominant region in the microscale and nanoscale strain-gradients.
In the drift-diffusion model, the drift-dominant region with respect to the temperature \textit{T} is given by 
\begin{equation} 
\begin{array}{cl}
k_{B}T < \cfrac{S(x){\nabla}u(x)}{{\nabla}S(x)}$\;$,
\end{array}
\label{eq2}
\end{equation}
where \textit{k$_B$}, \textit{S(x)}, and \textit{u(x)} are the Boltzmann constant, exciton generation rate under the Gaussian beam profile, and strain profile, respectively.
It indicates that the drift dominates over the diffusion when the right-hand term is larger than the thermal energy of $\sim$25 meV at room temperature \cite{harats2020}.
Fig.~\ref{fig:fig5}c shows the drift-dominant condition in our experimental condition. 
It shows the drift-dominant region of $\sim$100 nm, which occupies over $\sim$60 $\%$ of a whole strain-gradient region.
Hence, we achieve higher funnelling efficiency even with significantly smaller strain \textit{$\varepsilon$} $\approx$ 0.1 $\%$ compared to the microscale strain-gradient geometry to facilitate the low-threshold exciton transport. 
The increased exciton intensity with \textit{$\varepsilon$} $\approx$ 0.1 $\%$ in Fig.~\ref{fig:fig3}a is comparable to the microscale strain-gradient geometry with \textit{$\varepsilon$} $\approx$ 10 $\%$ \cite{lee2021}.
If we define a parameter $\alpha$ indicating a ratio between the funnelling efficiency and strain as 
\begin{equation} 
\begin{array}{cl}
\alpha = \cfrac{\eta}{\varepsilon}$\;$,
\end{array}
\label{eq3}
\end{equation}
the expected value is $\alpha$ $\approx$ 250 for a WSe$_2$ ML on the nano-gap (\textit{$\eta$} $\approx$ 25 $\%$, \textit{$\varepsilon$} $\approx$ 0.1 $\%$), which significantly exceeds $\alpha$ $\approx$ 1 of the previous study \cite{harats2020}.

The distribution of photo-excited excitons follows the illumination spot considering the area factor, as shown in Fig.~\ref{fig:fig5}d. 
The exciton density distribution is significantly modified by the strain-gradient due to the exciton funnelling effect. 
In addition, to include the trion conversion effect by the nano-gap, we estimate the electron density as a function of the distance based on the Boltzmann distribution as follows: 
\begin{equation} 
\begin{array}{cl}
n_{e}(x) = \cfrac{N_{0}e^{{\nabla}u_{c}(x)/k_{B}T}}{{\int}e^{{\nabla}u_{c}(x)/k_{B}T}xdx}$\;$,
\end{array}
\label{eq4}
\end{equation} 
where \textit{N$_0$} is the number of free carriers in the nano-gap and \textit{${\nabla}u_{c}\textit{(}x\textit{)}$} is the change of energy of the conduction band \cite{harats2020}.
From the estimated \textit{n$_e$(x)}, we obtain the distributions of exciton density and trion density for MoS$_2$ and WSe$_2$ MLs in our 
experimental condition, as shown in Fig.~\ref{fig:fig5}e and 5f (see Fig. S4 and S6 for more details).
Note that \textit{n$_{gap}$(x)} is a summation of \textit{n$_{X_0}$(x)} and \textit{n$_{X-}$(x)}.
In the simulated results, the formation of trion with decreased number of neutral exciton by the trion conversion is observed for a MoS$_2$ ML. 
In contrast, only increased number of neutral exciton is observed for a WSe$_2$ ML because \textit{n$_{e}\textit{(}x\textit{)}$} $\approx$ 0 for the p-doped WSe$_2$ ML \cite{moon2020}.
\\

\noindent
{\bf Discussion}

\noindent
Our work presents a low-threshold exciton transport and its dynamic control through the TMD on a nano-gap platform combined with TEPL spectroscopy. 
We experimentally observe $\sim$180 $\%$ increase of TEPL intensity with $\sim$0.1 $\%$ strain at the nano-gap, which is a comparable effect to the microscale strain-gradient induced with $\sim$10 $\%$ strain. 
That is, our approach allows a significant decrease of the strain threshold for facilitating recognizable exciton funnelling.
This result is attributed to the high funnelling efficiency of the nanoscale strain-gradient, which enables to occupy drift-dominant region over $\sim$60 $\%$ of the whole strain-gradient region.
In addition, spatial positioning with $\sim$0.2 nm precision and GPa scale pressure of the Au tip lead to the dynamic control of exciton dynamics, e.g., control of exciton funnelling rate in WSe$_2$ and trion conversion rate in MoS$_2$ in a reversible manner.
The presented efficient spatial manipulation of exciton dynamics at the nanoscale is highly desirable for further advances of exciton-based optoelectronics.
Thus, we envision that our approach will provide on-demand nanoscale positioning of the efficient strain-gradient platform enabling higher quantum efficiency and a greater degree of integration for the next generation exciton-based optoelectronic devices.
\\

\noindent
{\bf Methods}

\noindent
\textbf{Growth and transfer of WSe$_2$ MLs.}
In order to transfer commercially available chemical vapor deposition (CVD)-grown WSe$_2$ ML onto an nano-gap device, a wet transfer process was used. First, poly methyl methacrylate (PMMA) was spin-coated onto WSe$_2$ ML grown on the SiO$_2$ substrate. Then, the PMMA coated WSe$_2$ ML was separated from the SiO$_2$ substrate using a hydrogen fluoride solution, and carefully transferred onto the nano-gap device after being rinsed in distilled water to remove residual etchant. Next, it is dried naturally for 6 hours to improve the adhesion. Lastly, the PMMA was removed using acetone.
\\

\noindent
\textbf{Growth and transfer of MoS$_2$ MLs.}
A two-zone furnace was used to grow monolayer MoS$_2$ flakes. Sulfur flakes (Merck, $\ge$99.99 $\%$) were positioned in the upstream zone. As a molybdenum precursor, 0.01 M sodium molybdate aqueous solution was spun onto the SiO$_2$$/$Si substrate. The substrate was loaded into the downstream zone. The sulfur flakes and the substrate were heated with 200 $^{\circ}$C and 750 $^{\circ}$C temperatures for 7 min and maintained for 8 min. Then, the substrate was naturally cooled down to room temperature. The entire process was conducted under N$_2$ carrier gas with a flow rate of 600 sccm. Then, as-grown MoS$_2$ was coated with poly methyl methacrylate (PMMA) under 2500 rpm for 1 min. In order to delaminate SiO$_2$$/$Si substrate, the PMMA coated sample was floated on a 2 M KOH aqueous solution. After the delamination, underneath KOH residues were rinsed by deionized water several times. The PMMA/MoS$_2$ layer was scooped by the nano-gap patterned substrate. The PMMA layer was removed by acetone and isopropyl alcohol.
\\

\noindent
\textbf{Fabrication of the nano-gap by focused ion beam.}
Single-side polished silicon wafers with thermally grown SiO$_2$ with a thickness of 1 $\mu$m were purchased from University Wafers, Boston, USA. Electron-beam evaporation was used to deposit 50 nm gold on the wafers. FEI Nova 600 dual-beam system was used to perform focused ion beam (FIB) milling on the wafers to remove the gold from the silica surface creating the nano-gap. This was performed at an ion beam voltage of 30 kV and a current of 1 pA. 
\\

\noindent
\textbf{Fabrication of the nano-gap by atomic layer lithography.}
The rectangular shape nanogap was fabricated using atomic layer lithography technique \cite{chen2013}. First we performed a conventional photolithography with a negative photoresist of MA-N-1410(micro resist technology GmbH, Germany), which was spin-coated at 3000 rpm for 30 s onto a 500 $\mu$m thick silicon substrate, then soft-baked at 100 $^{\circ}$C for 90 s. The sample was exposed to a UV light (i-line, intensity about 10 mW$/$cm$^2$) for 40 s using a mask aligner (MA6, Suss Micro Tec, Garching, Germany). After the sample was developed by ma-D-533$/$S developer, Cr$/$Au layers in which their thickness are 5 nm and 100 nm, respectively, were deposited on patterned samples by an electron-beam evaporator, and a subsequent lift-off process is performed with an N-Methyl-2-pyrrolidone (NMP) solution.
A 20 nm-thick aluminum oxide (Al$_2$O$_3$) which determines the gap size was deposited by atomic layer deposition (ALD) at 250 $^{\circ}$C. After the ALD process, an 80 nm-thick Au layer was deposited on a patterned sample to fill inside rectangular holes, it was formed a Metal-Insulator-Metal(MIM) structure. A simple adhesive-tape-based peel-off process was performed in order to planarize the nanogap surface. Here, we used the Ar-ion miller (KVET-IM2000L, Korea Vacuum Tech, Gimpo, Korea) with an oblique angle of 85 $^{\circ}$, acceleration voltage of 80 V, beam current of 1.3 mA, and exposure time of 10 minutes, to make the peel-off process easier. Then, the sample was immersed in buffered oxide etchant 6$:$1 solution (J.T.Baker, United States) for 20 seconds to remove the top side of aluminum oxide layer (Al$_2$O$_3$). After these processes, the sample was exposed for 15 minutes under the oblique ion-milling conditions same as the above to smoothen any difference in the heights of the first and second metal layers.
\\

\noindent
\textbf{TEPL spectroscopy setup.}
The prepared TMD MLs on the nano-gap were loaded on a piezo-electric transducer (PZT, P-611.3X, Physik Instrumente) for XY scanning and applying the dynamic pressure control with $<$0.2 nm positioning precision. In order to press and release suspended TMD MLs, an Au tip (apex radius of $\sim$15 nm) was used. The Au tip, which was fabricated with a refined electrochemical etching protocol, was attached to a quartz tuning fork (resonance frequency of 32.768 kHz) to regulate the distance between the tip and sample based on shear-force AFM operated by a digital AFM controller (R9$+$, RHK Technology). For TEPL experiments, a conventional optical spectroscopy setup was combined with home-built shear-force AFM. For a high-quality wavefront of the excitation beam, a He-Ne laser (632.8 nm, $<$0.5 mW) was coupled and passed through a single-mode fiber (core diameter of $\sim$3.5 $\mu$m) and collimated again using an aspheric lens. The collimated beam was then passed through a half-wave plate to make the excitation polarization parallel to the tip axis. Finally, the beam was focused onto the Au tip using a microscope objective (NA $=$ 0.8, LMPLFLN100X, Olympus) with a side illumination geometry. In order to ensure highly efficient laser coupling to the Au tip, the tip position was controlled with $\sim$30 nm precision by Picomotor actuators (9062-XYZ-PPP-M, Newport). TEPL responses were collected using the same microscope objective (backscattering geometry) and passed through an edge filter (LP02-633RE-25, Semrock) to cutoff the fundamental laser line. TEPL signals were then dispersed onto a spectrometer (f $=$ 328 mm, Kymera 328i, Andor) and imaged with a thermoelectrically cooled charge-coupled device (CCD, iDus 420, Andor) to acquire TEPL spectra. Before the experiment, the spectrometer was calibrated with an Argon Mercury lamp. A 150 g$/$mm grating blazed to 800 nm (spectral resolution of 0.62 nm) was used for PL measurements.
\\

\noindent
\textbf{Simulation of the tip-induced local pressure, strain, force, and strain distribution.}
In order to quantify the tip-induced mechanical properties applied on the contact region between the Au tip and the TMD MLs, the experimental conditions were modeled and calculated using a commercially available 3D simulation program, the Mechanical Enterprise module from ANSYS. In the model, the material of the Au tip was set to gold in the program. The material properties of the TMD MLs, such as Young$’$s modulus, the Poisson ratio, and density, were derived from reference \cite{zhang2016}. The vertical position of the tip was located at the center of the nano-gap structure before starting the simulation. Both ends of the TMD MLs were considered to be not slipped on the Au substrate. In these conditions, the local pressure applied on the TMD MLs was calculated as a function of the pressing depth of the tip with a step of 0.2 nm.
\\

\noindent
\textbf{Data availability.}
The data that support the plots within this paper and other findings of this study are available from the corresponding author upon reasonable request.

\bibliography{nanogap} 

\vskip 1cm
\noindent
{\bf Acknowledgements}

\noindent
This work was supported by the 2018 Research Fund (1.180091.01) of UNIST(Ulsan National Institute of Science $\&$ Technology) and the National Research Foundation of Korea (NRF) grant funded by the Korea government (MEST) (2019K2A9A1A06099937 and 2020R1C1C1011301).
H.-R. Park acknowledges NRF-2021R1A2C1008452.
S.H. Choi and K.K. Kim acknowledge the support by the Institute for Basic Science (IBS-R011-D1).
K.K. Kim acknowledges the Basic Research Program through the National Research Foundation of Korea (NRF) funded by the Ministry of Science, ICT $\&$ Future Planning (2018R1A2B2002302)

\noindent
{\bf Author contributions}

\noindent
H. Lee, Y. Koo, and K.-D Park conceived the experiments.
H. Lee and Y. Koo performed the TEPL spectroscopy and control experiments.
S. Kumar, H.-T. Lee, G. Ji, H.-R. Park, and H. Choo designed and fabricate the nano-gap.
S.H. Choi and K.K. Kim prepared and transfered TMD MLs on the Au nano-gap device.
J. Choi performed theoretical calculation and modelling of exciton distribution.
H. Lee, Y. Koo, J. Choi, M. Kang, and K.-D. Park analyzed the data, and all authors discussed the results.
H. Lee Y. Koo, and K.-D. Park wrote the manuscript with contributions from all authors.
K.-D. Park supervised the project.

\noindent
{\bf Additional information}

\noindent
Supplementary information is available for this paper.

\noindent
{\bf Competing financial interests}

\noindent
The authors declare no competing financial interests.

\end{document}